\documentclass[twocolumn,showpacs,floatfix]{revtex4}

\usepackage{color}
\usepackage{ulem}
\usepackage{graphicx}
\usepackage{dcolumn}
\usepackage{bm}
\usepackage{here}

\begin{document}

\preprint{APS/123-QED}

\title{
Magnetic-field- and pressure-induced quantum phase transition in CsFeCl$_3$
proved via magnetization measurements
}

\author{Nobuyuki Kurita and Hidekazu Tanaka}

\affiliation{
Department of Physics, Tokyo Institute of Technology, Meguro-ku, Tokyo 152-8551, Japan
}

\date{\today}

\begin{abstract}
We have performed magnetization measurements of the gapped quantum magnet CsFeCl$_3$
at temperatures ($T$) down to 0.5\,K at ambient pressure
and down to 1.8\,K at hydrostatic pressures ($P$) of up to 1.5\,GPa.
The lower-field ($H$) phase boundary of the 
field-induced ordered phase at ambient pressure is found to follow the power-law behavior expressed by the formula $H_{\rm N}(T)$\,$-$\,$H_{\rm c}$\,$\propto$\,$T_{\rm N}^{\phi}$.
The application of pressure extends the phase boundary
to both a lower field and higher temperature. 
Above the critical pressure $P_{\rm c}$\,$\sim$\,0.9\,GPa, 
the transition field $H_{\rm N}$ associated with the excitation gap becomes zero, 
and a signature of the magnetic phase transition is found in 
the $T$-dependence of magnetization in a very low applied field.
This suggests that CsFeCl$_3$ exhibits a pressure-induced magnetic phase 
transition at $P_{\rm c}$. 
\end{abstract}


\maketitle
\section{Introduction}
Over the past decade, gapped quantum spin systems 
have attracted considerable attention owing to their variety of phase transitions.
The gapped ground state is typically a spin-singlet 
with an excitation gap $\Delta$ to the lowest excited state.
The complete suppression of $\Delta$ by varying external parameters often triggers a phase transition.
In particular, a quantum phase transition (QPT) which is a continuous phase transition occurring at zero temperature ($T$) 
as a consequence of quantum fluctuations is the fundamental subject
that correlates condensed matter physics with particle physics\,\cite{Hertz_PRB1976,Sachdev_Cambridge1999}.
It is well established that the transition point, 
commonly known as the quantum critical point (QCP),
can be accessed continuously by applying a magnetic field ($H$) and/or hydrostatic pressure ($P$).
The field-induced QPT differs from the pressure-induced QPT in terms of the universality class of the QCP.
For the former and latter QPTs, magnetic excitations have quadratic and linear dispersion relations, respectively,
at the QCP, where the excited mode becomes gapless.

Recent intensive studies have shown that a field-induced QPT to an XY antiferromagnetic (AF) phase
can be described in the context of the Bose--Einstein condensation (BEC) 
of magnon quasiparticles\,\cite{Affleck_PRB1991,Giamarchi_PRB1999,Nikuni_PRL2000}.
The uniaxial symmetry around the applied field in the spin Hamiltonian [$O$(2)] is effectively
translated into the conservation of the total number of particles [$U$(1)].
This concept is useful for understanding a QPT from the standpoint of a dilute Bose gas system.
Experimentally, the magnon BEC scenario has been examined using several gapped quantum magnets\,\cite{Oosawa_JPCM1999,Oosawa_PRB2001,TanakaH_JPSJ2001,Ruegg_Nature2003,YamadaF_JPSJ2008,Sagao_PRB1997,Jaime_PRL2004,Sebastian_PRB2005,Sebastian_Nature2006,Mazurenko_PRL2014,Honda_JPCM1997,Tsujii_PRB2005,Shiramura_JPSJ1997,Paduan-Filho_PRB2004,Zapf_PRL2006}.
However, the focus has mostly been on systems with weakly coupled spin dimers such as TlCuCl$_3$\,\cite{Oosawa_JPCM1999,Oosawa_PRB2001,TanakaH_JPSJ2001,Ruegg_Nature2003,YamadaF_JPSJ2008} and 
BaCuSi$_2$O$_6$\,\cite{Sagao_PRB1997,Jaime_PRL2004,Sebastian_PRB2005,Sebastian_Nature2006,Mazurenko_PRL2014}.

A pressure-induced QPT in quantum spin systems is also of importance,
particularly because it provides a rare opportunity 
to directly identify the massive Higgs mode separately from the massless Nambu--Goldstone mode.
The Higgs mode is a collective mode of amplitude oscillations of order parameters\,\cite{Sachdev_Cambridge1999,Matsumoto_PRB2004,Matsumoto_JPSJ2007,Pekker_ACMP2015}.
A requirement for observing the Higgs mode is shrinkage of the ordered moment
in zero applied field, as realized in the pressure-induced ordered phase.
Thus far, to our knowledge, TlCuCl$_3$\,\cite{Oosawa_JPSJ2004,Goto_JPSJ2004,Ruegg_PRL2008} 
and KCuCl$_3$\,\cite{Goto_JPSJ2006,Goto_JPSJ2007} are the only quantum magnets
for which a pressure-induced QPT to the ordered phase has been established. 
Recently, neutron scattering experiments on TlCuCl$_3$ have provided evidence for the Higgs mode\,\cite{Ruegg_PRL2008,Merchant_NatPhys2014}. 

\begin{figure}[b]
\begin{center}
\includegraphics[width=0.9\linewidth]{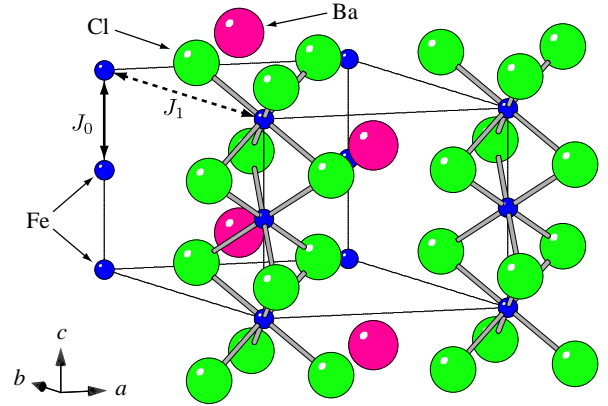}
\end{center}
\caption{(Color online) 
Crystal structure of CsFeCl$_3$. Thin solid lines denote the chemical unit cell. 
Double-headed solid and dashed arrows are the intrachain and 
interchain exchange interactions $J_0$ and $J_1$, respectively.
}
\label{fig1}
\end{figure}

The compound CsFeCl$_3$ crystallizes in a hexagonal structure, 
as shown in Fig.~\ref{fig1}\,~\cite{Seifert_ZAAC1966},
in which magnetic Fe$^{2+}$ ions are surrounded octahedrally by six Cl$^-$ ions. 
In CsFeCl$_3$, magnetic Fe$^{2+}$ ions that align along the $c$-axis 
form ferromagnetic (FM) spin chains\,~\cite{Steiner_SSC1981}. 
These FM spin chains form a regular triangular lattice in the basal $ab$-plane
with weak AF interactions, as shown in Fig.~\ref{fig1}.

The low-temperature (LT) magnetic moment of Fe$^{2+}$ in an octahedral environment is
 determined by the lowest orbital triplet $T_{2g}$\,~\cite{Oguchi_JPSJ1967}. 
This orbital triplet splits into three singlet and six doublets owing to spin-orbit coupling 
and the uniaxial crystal field, which are expressed together as
\begin{eqnarray}
{\cal H}^{\prime}=-k{\lambda}({\bm l}\cdot{\bm S})-{\delta}\left\{(l^z)^2-2/3\right\},
\label{eq:perturb}
\end{eqnarray}
where $\bm l$ is the effective angular momentum with $l\,{=}\,1$, 
$\bm s$ is the true spin with $S\,{=}\,2$, and $k\ ({\sim}\,0.9)$ is the reduction factor, 
which expresses the fact that the matrix elements of the angular momentum $\bm l$ are 
reduced owing to the mixing of the $p$ orbitals of the surrounding Cl$^-$ with the $3d$ orbitals of Fe$^{2+}$. 
When the temperature $T$ is much lower than the magnitude of the spin-orbit coupling constant 
${\lambda}\,{\simeq}\,{-}\,100$ cm$^{-1}$, i.e., $T\,{\ll}\, |{\lambda}|/k_{\rm B}\,{\simeq}\,150$\,K, 
the magnetic property is determined by the lowest singlet and doublet, 
which are given by $m\,{=}\,0$ and ${\pm}1$, respectively, with $m\,{=}\,l^z\,{+}\,S^z$. 
When the FeCl$_6$ octahedron is trigonally elongated, as observed in CsFeCl$_3$, 
the energy of the $m\,{=}\,0$ state is lower than that of the $m\,{=}\,{\pm}1$ state. 
Hence, using the effective spin $s\,{=}\,1$, the LT magnetic properties of CsFeCl$_3$ 
can be described by the Hamiltonian\,~\cite{Oguchi_JPSJ1967}
\begin{eqnarray}
{\cal H}&=&\sum_i D\left(s_i^z\right)^2-\sum_{\langle i,j\rangle}^{\rm chain} J_0\left(s_i^xs_j^x+s_i^ys_j^y+{\Delta}s_i^zs_j^z\right)\nonumber\\
&+&\sum_{\langle l,m\rangle}^{\rm plane} J_1\left(s_l^xs_m^x+s_l^ys_m^y+{\Delta}s_l^zs_m^z\right),
\end{eqnarray}
where the first term is the single-ion anisotropy ($D\,{>}\,0$) corresponding to the energy difference 
between the $m\,{=}\,{\pm}1$ and $m\,{=}\,0$ states, and the second and third terms are the FM exchange interaction 
in the chain and the AF exchange interaction in the $ab$-plane, respectively. 
$\Delta$ is the exchange anisotropy. 
The coupling constants determined from the dispersion relations are $D/k_{\rm B}\,{=}\,25.3$\,K, 
$J_0/k_{\rm B}\,{=}\,5.27$\,K, and $J_1/k_{\rm B}\,{=}\,0.28$\,K\,~\cite{Yoshizawa_JPSJ1980}. 
The anisotropy parameter $\Delta$ is expected to be $0\,{<}\,{\Delta}\,{<}\,1$ 
because $D\,{>}\,0$\,~\cite{Oguchi_JPSJ1967}. 
However, its value is unclear.

CsFeCl$_3$ has a gapped ground state~\cite{Yoshizawa_JPSJ1980,Steiner_SSC1981,Knop_JMMM1983} and exhibits an AF ordering when a magnetic field is applied 
along the hexagonal $c$-axis ($H$\,$\parallel$\,$c$)\,\cite{Haseda_PhysicaB1981}.
Unlike the case of the spin dimer system, the gapped ground state originates from
competition between the large easy-plane single ion anisotropy $D(s^z)^2$ and the exchange interactions.

Field-induced AF ordering with $H\,{\parallel}\,c$ has been confirmed in several experiments
in fields between $\sim$\,4\,T and \,$\sim$\,11\,T at LTs below 2.6\,K\,\cite{Haseda_PhysicaB1981,Baines_JPC1983,Tsuboi_JPCC1988,Chiba_JPC1988,Toda_PRB2005}.
Neutron scattering experiments have revealed that the ground-state spin configuration is probably
a 120$^\circ$ structure with the wave vector $q$\,$\approx$\,(1/3,1/3,0)
characteristic of triangular-lattice antiferromagnets\,\cite{Toda_PRB2005}. 
The order parameter has been deduced to be perpendicular magnetization $M_{xy}$
from the temperature and field variations of $M_{xy}$, which appears to be
in accordance with magnon BEC theory\,\cite{Toda_PRB2005,Tsuneto_Physica1971}.
A useful feature for experimentally characterizing the magnon BEC is the power-law behavior 
for the phase boundary in the vicinity of  $T$=0 and the cusplike minimum of the magnetization
at the ordering temperature $T_{\rm N}(H)$.
Thus far, however, the magnetic phase diagram of CsFeCl$_3$ has not been established sufficiently.
Interestingly, a preliminary high-pressure magnetization study on this compound
suggests the occurrence of a pressure-induced QPT\,\cite{Sasaki_PTP2005}.

In this paper, we present the results of magnetization measurements of CsFeCl$_3$
a temperatures down to 0.5\,K at ambient pressure and down to 1.8\,K at high pressures.
The power law behavior for the lower-field phase boundary of the field-induced ordered phase
at ambient pressure is discussed.
With increasing pressure, the ordered phase extends systematically towards 
both a lower-field and higher-temperature.
It is found that with increasing pressure, the excitation gap $\Delta$ decreases systematically 
and appears to be zero at $P_{\rm c}$\,$\sim$\,0.9\,GPa.
For  $P$\,$\ge$\,$P_{\rm c}$, magnetic ordering emerges in the a very low magnetic field.

\section{Experimental Details}

Single crystals of CsFeCl$_3$ were grown via the vertical Bridgman method 
from a melt comprising a stoichiometric mixture of CsCl and FeCl$_2$ sealed in an evacuated quartz tube.
The ingredients were dehydrated in vacuum by heating at 80\,--\,150$^\circ$C for three days. 
The temperature at the center of the furnace was set at 640$^\circ$C and 
the crystals were lowered at a rate of 3\,mm/hour. 
We repeated the same procedure after the removal of impurities and imperfect crystals.
The single crystals obtained were confirmed to be CsFeCl$_3$ by X-ray diffraction.

The magnetization was measured down to 1.8\,K under magnetic fields of up to 7\,T 
parallel to the $c$-axis using a SQUID magnetometer (MPMS-XL, Quantum Design).
At ambient pressure, a $^3$He system (iHelium3, IQUANTUM) was 
used for the measurement down to the lowest temperature $T_{\rm min}$ of 0.5\,K.

Magnetization measurements under hydrostatic pressure were performed up to
a pressure of 1.5\,GPa using a clamped piston cylinder pressure device.
Daphne\,7373 (Idemitsu Kosan), which remains in the liquid state 
up to $\sim 2$\,GPa at room temperature\,\cite{Daphne7373},
was used as a pressure-transmitting medium.
The pressure generated in the sample space was calibrated at a low temperature 
by the change in the superconducting transition temperature $T_{\rm c}$ of tin under $H=10$\,Oe.
The narrow transition width remains almost unchanged up to the maximum pressure 
of 1.5\,GPa, indicating that the nonhydrostatic effect is negligibly small.
The high-pressure magnetization data presented in this paper were corrected 
to remove the background contribution of the pressure device\,\cite{MPMS_BG}.
For the high-pressure experiments, we used three pieces of single crystals from different batches and
confirmed no obvious sample dependence.

\section{Results and Discussion}

\subsection{Ambient pressure magnetization}

\begin{figure}
\begin{center}
\includegraphics[width=0.9\linewidth]{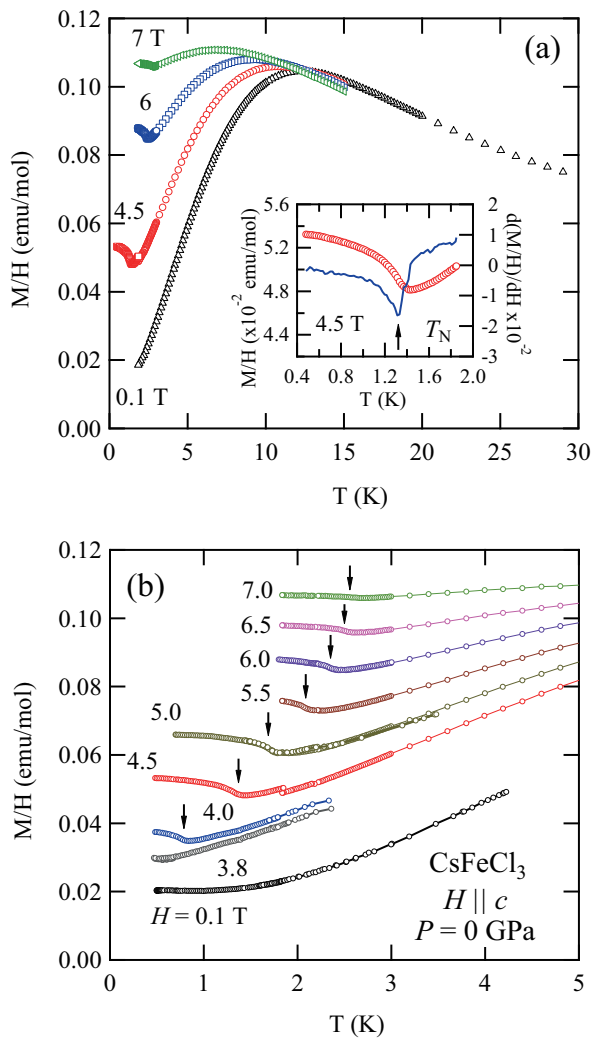}
\end{center}
\caption{(Color online) 
(a) Magnetic susceptibility $\chi$(=$M/H$) vs $T$ for CsFeCl$_3$ at ambient pressure
under several fields of up to 7\,T for $H \parallel c$.
(b) Low-temperature expanded view of the $\chi(T)$ data. 
Arrows indicate the transition temperature $T_{\rm N}$, which is defined as the temperature 
with the peak in d$\chi$/d$T$($T$) as shown in the inset of (a). 
} \label{fig2}
\end{figure}

Figure~\ref{fig2}(a) shows the temperature dependence of the magnetic susceptibility $\chi$(=$M/H$) 
of CsFeCl$_3$ at ambient pressure under several fields of up to 7\,T for $H$\,$\parallel$\,$c$.
For the 0.1\,T data, $\chi(T)$ exhibits a broad maximum at approximately 12\,K 
with decreasing temperature, followed by a rapid decrease toward zero.
With increasing field, the broad maximum shift to lower temperatures
while $\chi(T)$ at LTs increases.
As shown in Fig.~\ref{fig2}(b), $\chi(T)$ decreases monotonically
down to $T_{\rm min}$\,=\,0.5\,K under fields of up to 3.8\,T,
indicative of the gapped ground state up to 3.8\,T.
The finite magnetic susceptibility of ${\chi}_{\rm VV}\,{\simeq}\,0.02$ emu/mol below 1\,K for $H\,{=}\,0.1$\,T 
is attributed to the large temperature-independent Van Vleck paramagnetism of Fe$^{2+}$ 
in the octahedral environment, as in the case of Co$^{2+}$\,~\cite{Susuki_PRL2013}.
At higher fields of above 3.8\,T, a magnetic phase transition appears as a cusplike minimum in $\chi(T)$,
which is a characteristic of magnon BEC\,~\cite{Nikuni_PRL2000}.
We assign the transition temperature $T_{\rm N}$ as the temperature with the peak 
in d$\chi$/d$T$($T$) as displayed in the inset of Fig.~\ref{fig2}(a).
With increasing magnetic field, $T_{\rm N}$ increases as indicated by arrows.
These results are consistent with previous reports\,\cite{Haseda_PhysicaB1981,Baines_JPC1983,Tsuboi_JPCC1988,Chiba_JPC1988,Toda_PRB2005,Sasaki_PTP2005}
except that we could only detect a single phase transition
instead of three successive phase transitions observed in a previous specific heat study\,\cite{Haseda_PhysicaB1981}.
We do not yet have a plausible explanation for the difference.
Note that a single phase transition was also observed in our specific heat measurements,
and that the $T_{\rm N}$ values obtained by two different methods in our studies are 
consistent with each other\,\cite{Unpublished}.

\begin{figure}
\begin{center}
\includegraphics[width=0.9\linewidth]{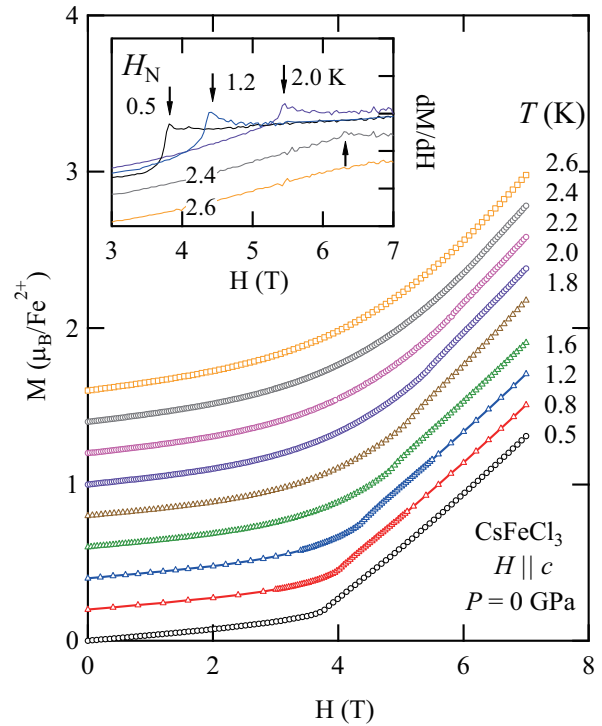}
\end{center}
\caption{(Color online) 
Magnetization curves $M(H)$ of CsFeCl$_3$ for $H$\,$\parallel$\,$c$ 
at ambient pressure and several temperatures.
The $M(H)$ data are shifted in the longitudinal direction by 0.2\,${\rm \mu_B}$/Fe$^{2+}$ 
with each increase in temperature for clarity.
The inset shows d$M/$d$H(H)$  vs $H$ at selected temperatures.
Arrows indicate the transition field $H_{\rm N}$.
} \label{fig3}
\end{figure}

Figure~\ref{fig3} shows the field dependence of the magnetization $M(H)$ of CsFeCl$_3$ 
for $H$\,$\parallel$\,$c$ at ambient pressure and several temperatures.
The $M(H)$ data are shifted in the longitudinal direction 
by 0.2\,${\rm \mu_B}$/Fe$^{2+}$ with each increase in temperature for clarity.
At 0.5\,K, $M(H)$ exhibits a clear kink-like anomaly at approximately 4\,T,
which corresponds to a phase transition from the gapped state to the AF ground state.
This anomaly is more clearly observed in the field derivative d$M/$d$H(H)$ 
as shown in the inset of Fig.~\ref{fig3}.
The transition field $H_{\rm N}$ is defined as the field 
where d$M/$d$H(H)$  shows a peak- or shoulder-like anomaly.
With increasing temperature, $H_{\rm N}$ increases 
while the anomaly becomes broadened and is no longer detectable at 2.6\,K.
Below $H_{\rm N}$, $M(H)$ exhibits a continuous increase 
in spite of the gapped nonmagnetic ground state.
This is mostly attributed to the large Van Vleck paramagnetism 
arising from the crystal field effect.
A similar feature has also been found in the isomorphic compound CsFeBr$_3$,
where the ground state is gapped\,\cite{TanakaY_JPSJ2001}.

\begin{figure}
\begin{center}
\includegraphics[width=0.9\linewidth]{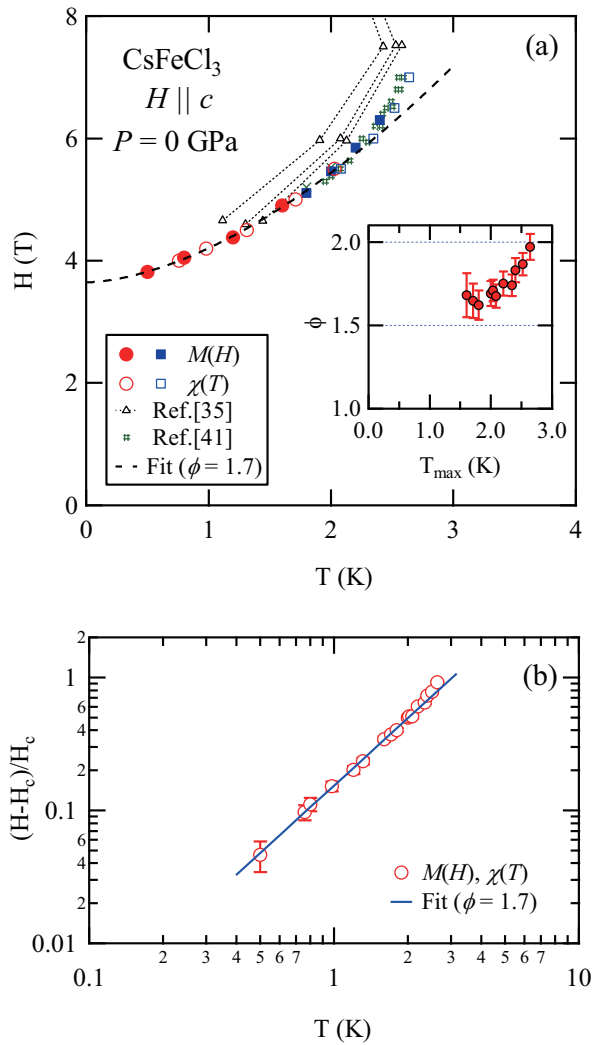}
\end{center}
\caption{(Color online) 
(a) $H$\,$-$\,$T$ phase diagram of CsFeCl$_3$ for $H$\,$\parallel$\,$c$ 
at ambient pressure determined from magnetization measurements.
Circle and squares were obtained using MPMS and iHelium3 instruments, respectively.
Open and solid symbols were determined from $\chi(T)$ and $M$($H$) data, respectively.
The dashed curve is a fit to the power law $H-H_{\rm N}$\,$\propto$\,$T_{\rm N}^{\phi}$ 
with $\phi$\,=\,1.7.
For comparison, data from Refs.\,\onlinecite{Haseda_PhysicaB1981} and \onlinecite{Sasaki_PTP2005}
are also plotted.
The inset shows $\phi$ vs $T_{\rm max}$, where
$\phi$ was evaluated using data between 0.5\,K (fixed) and several values of $T_{\rm max}$.
(b)  Double logarithmic plot of the reduced field $(H-H_{\rm c})$/$H_{\rm c}$ against $T$.
The solid line is a fit with $\phi$\,=\,1.7.
} \label{fig4}
\end{figure}

In Fig.~\ref{fig4}(a), we illustrate the $H$\,$-$\,$T$ phase diagram of CsFeCl$_3$ 
for $H$\,$\parallel$\,$c$ at ambient pressure, 
determined from magnetization measurements down to 0.5\,K.
The high-temperature data are in good agreement 
with the results of a previous magnetization study\,~\cite{Sasaki_PTP2005},
while three successive phase transitions were reported in Ref.\,\onlinecite{Haseda_PhysicaB1981}.
The dashed curve represents a fit to data using  
the power law $H_{\rm N}(T)$\,$-$\,$H_{\rm c}$\,$\propto$\,$T_{\rm N}^{\phi}$ with $\phi$\,=\,1.7.
The power law behavior at low temperatures is more clearly observed in the double logarithmic plot of the reduced field $(H-H_{\rm c})$/$H_{\rm c}$ against $T$, as shown in Fig.~\ref{fig4}(b).
The solid line is a fit to data with $\phi$\,=\,1.7.
This power law assumes a dilute boson limit and 
hence is only valid at sufficiently low temperatures
as compared with the energy scale of boson interactions or AF couplings.
We evaluated the exponent $\phi$ from a best fit with the power law to the data between $T_{\rm min}$\,=\,0.5\,K (fixed) 
and various temperatures $T_{\rm max}$ ranging from 1.6\,K to 2.6\,K.
The critical field $H_{\rm c}$ ($H_{\rm N}$ at $T$\,$=$\,0), which is set to be free during fitting, 
was obtained as 3.6\,--\,3.7\,T.
The inset of Fig.~\ref{fig4}(a) shows $\phi$ as a function of $T_{\rm max}$.
With decreasing $T_{\rm max}$, $\phi$ decreases and tends 
to approach ${\phi}_{\rm BEC}\,{=}\,1.5$, the critical exponent predicted for three-dimensional BEC\,~\cite{Affleck_PRB1991,Giamarchi_PRB1999,Nikuni_PRL2000},
rather than the value of 1.0 for two-dimensional BEC\,~\cite{Batista_PRL2007}.
The overestimate of $\phi$ compared with ${\phi}_{\rm BEC}\,{=}\,1.5$ in this study is probably because 
the temperature range employed for the analysis was not sufficiently low.
This is supported by theoretical calculations demonstrating that, 
as the analyzed temperature range is reduced,
$\phi$ decreases and converges at ${\phi}_{\rm BEC}\,{=}\,1.5$\,~\cite{Nohadani_PRB2004,Nohadani_PRB2005,Kawashima_JPSJ2004,Kawashima_JPSJS2005}.
A similar feature has also been found in several quantum spin systems.
In TlCuCl$_3$, for instance, $\phi$\,=\,2.0\,--\,2.2, obtained 
at temperatures of above 1.8\,K\,~\cite{Oosawa_PRB2001,TanakaH_JPSJ2001},
was refined to 1.67 when the measurement was performed 
down to the lower temperature of 0.5\,K\,~\cite{Shindo_JPSJ2004}.
The $\phi$ value eventually converged to 1.5 
according to the results of magnetization measurement down to 77\,mK\,~\cite{YamadaF_JPSJ2008}.
Note that $\phi$\,=\,1.6\,--\,1.7 obtained for CsFeCl$_3$ in this study is consistent 
with the case of TlCuCl$_3$ using a similar lowest temperature.
To more accurately determine the critical exponent for CsFeCl$_3$,
further experiments at lower temperatures are required.

\subsection{High-pressure magnetization}

\begin{figure}
\begin{center}
\includegraphics[width=0.9\linewidth]{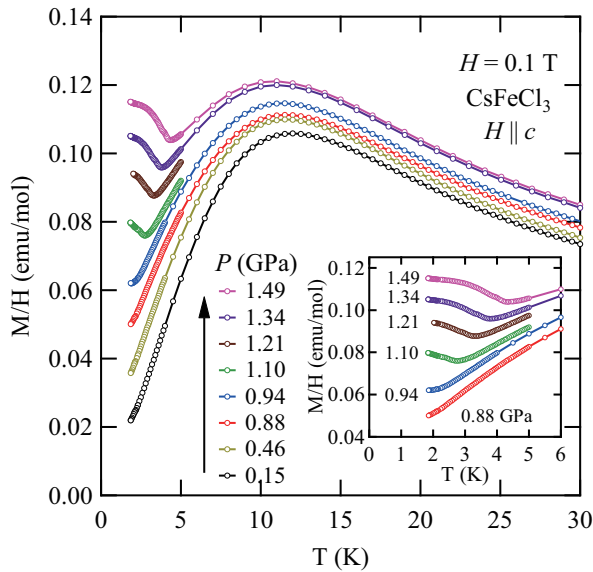}
\end{center}
\caption{(Color online) 
(a) $M/H$ vs $T$ for CsFeCl$_3$ with $H\parallel c$ under 0.1\,T
at several pressures.
The inset shows an expanded view focusing on high pressure data at low temperatures.
} \label{fig5}
\end{figure}

Figure~\ref{fig5} shows the magnetic susceptibility $\chi$ (=$M/H$) vs $T$ for CsFeCl$_3$ 
under 0.1\,T with $H$\,$\parallel$\,$c$ at several pressures.
No significant changes can be observed in the overall features of $\chi(T)$ up to 0.88\,GPa.
This indicates that the ground state remains gapped at $P$\,$\le$\,0.88\,GPa.
At high pressures of above 0.94\,GPa, $\chi(T)$ at LTs exhibits minima
that shift to higher temperatures with increasing pressure.
The minimum of $\chi(T)$ is attributable to a magnetic phase transition
because this behavior is similar to $\chi(T)$ at ambient pressure for $H\,{>}\,H_{\rm c}$, 
as shown in Fig.~\ref{fig2}(b)
Thus, the critical pressure $P_{\rm c}$, where the excitation gap is closed and
the ordered ground state appears, is evaluated to be $\sim$\,0.9\,GPa under 0.1\,T.
No obvious anomalies related to $T_{\rm N}$ can be observed under 0.1\,T,
as opposed to the sharp peaks in d$\chi$/d$T$ under higher fields.
We hence assign the temperature exhibiting the peak in d$\chi$/d$T$ as
the transition temperature $T_{\rm N}$(0.1\,T) for the pressure-induced ordered phase.

\begin{figure}
\begin{center}
\includegraphics[width=0.95\linewidth]{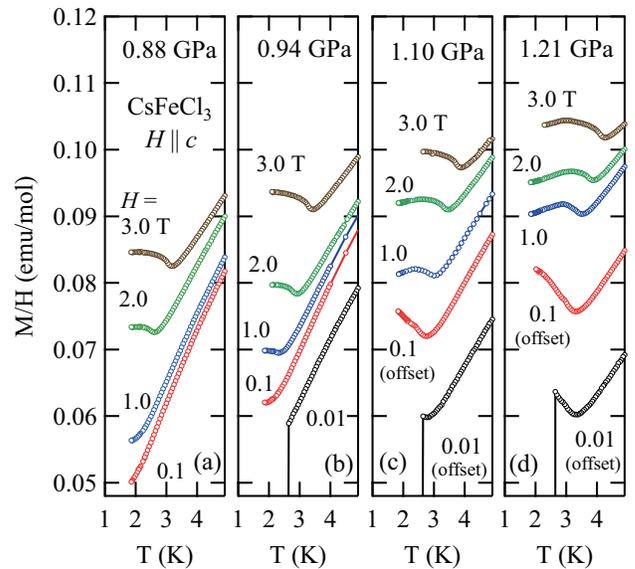}
\end{center}
\caption{(Color online) 
(a) $M/H$ vs $T$ for CsFeCl$_3$ with $H\parallel c$ under several fields
at $P=$ (a) 0.88\,GPa, (b) 0.94\,GPa, (c) 1.10\,GPa, and (d) 1.21\,GPa.
The $M/H$($T$) data under low fields are shifted in the longitudinal direction for clarity.
} \label{fig6}
\end{figure}

Figure~\ref{fig6} shows the $\chi(T)$ data under several fields at 
$P$\,=\,0.88\,GPa, 0.94\,GPa, 1.10\,GPa, and 1.21\,GPa.
The discontinuous behavior observed for the 0.01\,T data below $\sim$\,3\,K is caused by  
the Meissner effect induced by a superconducting transition of tin, which is included in 
the sample space as a pressure manometer.
$T_{\rm N}$ for $H$\,$\ge$\,0.1\,T systematically increases with increasing pressure.
Note that, as shown in Figs.~\ref{fig6}(c) and (d), 
the minimum of $\chi$($T$) indicative of $T_{\rm N}$ can also be found 
in the very low field of 0.01\,T.
In addition, $T_{\rm N}$ under 0.01\,T appears to increase 
with increasing pressure, similarly to the higher field data.
Consequently, we can deduce that the magnetic phase transition occurs 
in zero magnetic field at 1.10\,GPa.

\begin{figure}
\begin{center}
\includegraphics[width=0.95\linewidth]{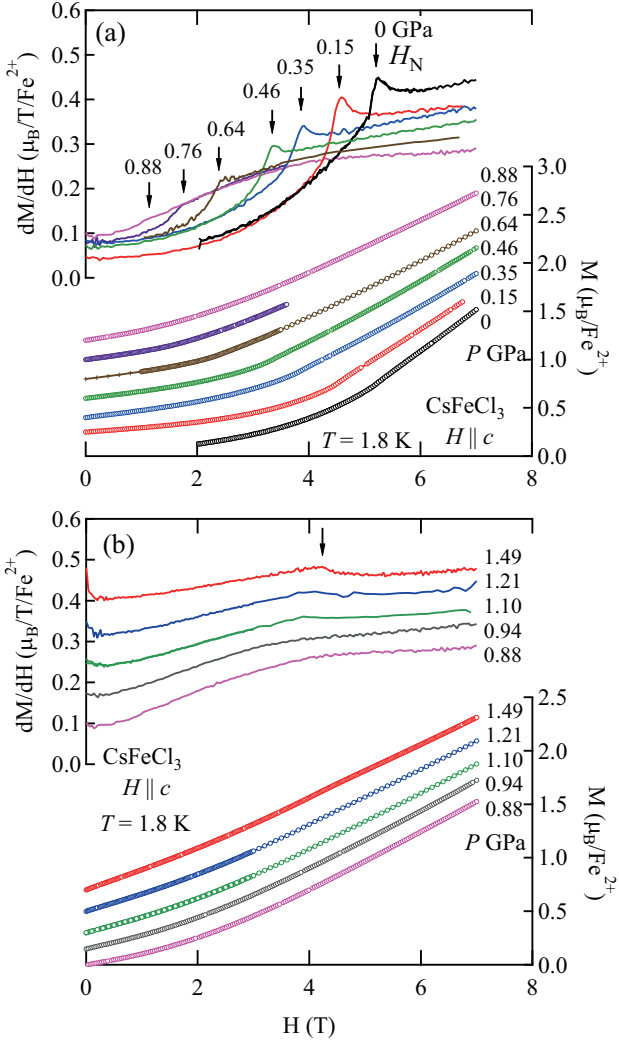}
\end{center}
\caption{(Color online) 
Field dependence of $M$ (left) and d$M$/d$H$ (right) 
for CsFeCl$_3$ with $H$\,$\parallel$\,$c$
at 1.8\,K for (a) $P$\,$\le$\,0.88\,GPa and (b)  0.88\,$\le$\,$P$\,$\le$\,1.49\,GPa.
The data except for d$M$/d$H(H)$ in (a) are shifted arbitrarily in the longitudinal direction for clarity.
Arrows in (a) and (b) indicate $H_{\rm N}$ and a possible change in the spin structure, respectively.
} \label{fig7}
\end{figure}

Figures~\ref{fig7}(a) and (b) show the $M(H)$ and d$M$/d$H(H)$ data for CsFeCl$_3$ 
with $H$\,$\parallel$\,$c$ at 1.8\,K under several pressures of up to 1.5\,GPa.
With increasing pressure, $H_{\rm N}$, defined by a peak or shoulder in d$M$/d$H(H)$,
decreases although the peak becomes smeared [Fig.~\ref{fig7}(a)].
No obvious anomaly related to $H_{\rm N}$ exists above 0.88\,GPa [Fig.~\ref{fig7}(b)].
Since $H_{\rm N}$ is a measure of the $\Delta$ value, 
the present results indicate that the application of pressure continuously decreases $\Delta$ 
up to 0.88\,GPa, above which the excitation gap is completely closed 
and the ground state is an AF ordered state.
Note that, as indicated by an arrow in Fig.~\ref{fig7}(b), 
the shoulder-like behavior in d$M$/d$H(H)$ for high-pressure data around 4\,T   
evolves to a cusplike peak with increasing pressure.
The cusplike peak suggests a change in the spin structure in the $ab$-plane, 
because spins are forced to lie in the $ab$-plane owing to the large $D$ term.
A spin reorientation transition under hydrostatic pressure has been observed in TlCuCl$_3$\,\cite{Oosawa_JPSJ2004,Yamada_PRB2008}. 
This transition was interpreted to result from the fourth-order anisotropy, 
which becomes effective when the magnitude of the moment is large.
To clarify the magnetic-field induced transition in CsFeCl$_3$ at high pressures, 
further experiments are necessary.

\begin{figure}
\begin{center}
\includegraphics[width=0.95\linewidth]{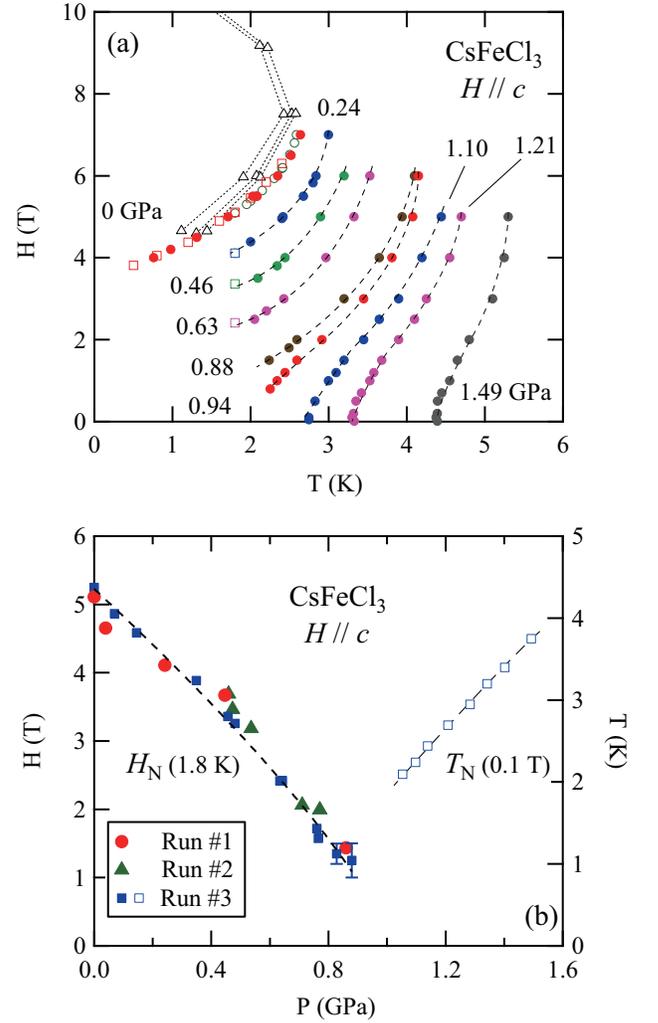}
\end{center}
\caption{(Color online) 
(a) Pressure evolution of the $H$\,$-$\,$T$ phase diagram of CsFeCl$_3$ 
for $H \parallel c$, determined via magnetization measurements.
Dashed curves are guides to the eyes.
(b) $H_{\rm N}$($T$\,$=$\,1.8\,K) and $T_{\rm N}$($H$\,=\,0.1\,T) as a function of pressure.
The dashed curves for $H_{\rm N}$ and $T_{\rm N}$ 
are a fit (see text) and a guide to the eyes, respectively.
} \label{fig8}
\end{figure}

Figure~\ref{fig8}(a) shows the pressure evolution of the lower-field boundary of 
the ordered phase in CsFeCl$_3$ for $H$\,$\parallel$\,$c$ at selected pressures, 
determined via magnetization measurements.
With increasing pressure, the phase boundary systematically 
moves toward a higher-temperature and lower-field.
At $P_{\rm c}$\,$\sim$\,0.9\,GPa, the zero-field ground state 
changes from a gapped state to a magnetically ordered state.
Note that the $H$\,$-$\,$T$ phase diagram of CsFeCl$_3$ 
in the pressure-induced ordered phase at $P$\,$\ge$\,$P_{\rm c}$ resembles that of 
the isomorphic compound RbFeCl$_3$ at ambient pressure\,\cite{Haseda_PhysicaB1981}.
RbFeCl$_3$ exhibits magnetic ordering in zero field
owing to the relatively large exchange interactions as compared with the $D$ term.
Figure~\ref{fig8}(b) shows the pressure dependence of 
$H_{\rm N}$($T$\,$=$\,1.8\,K) and $T_{\rm N}$($H$\,=\,0.1\,T).
The dashed curve for the $H_{\rm N}$ ($\sim$\,$\Delta$) data is a fit using the empirical formula 
($P$\,$-$\,$P_{\rm c}$)$^{\alpha}$\,$\propto$\,$\Delta$.
The exponent $\alpha$ was obtained to be 0.77.
For comparison, $\alpha$\,=\,0.33 has been reported for TlCuCl$_3$\,~\cite{Goto_JPSJ2004}.
It is noted that, as found at ambient pressure, the transition field $H_{\rm N}$ evaluated at 0.5\,K
is smaller than that at the lowest investigated temperature of 1.8\,K.
This is probably the main reason why the two parameters in Fig.~\ref{fig8}(b)
appear not to converge a single critical point.
Lower-temperature measurements are necessary to more precisely determine
the value of $P_{\rm c}$.

The relationship between the excitation gap and exchange interactions in CsFeCl$_3$ can be derived as 
\begin{eqnarray}
\Delta = \sqrt{D^2-2D(2J_0+3J_1)},
\label{gap}
\end{eqnarray} 
within the mean field theory\,\cite{TanakaY_JPSJ2001}.
Using $D/k_{\rm B}\,{=}\,25.3$\,K, $J_0/k_{\rm B}\,{=}\,5.27$\,K, 
and $J_1/k_{\rm B}\,{=}\,0.28$\,K determined at ambient pressure\,~\cite{Yoshizawa_JPSJ1980} 
and $g_{\parallel}\,{=}\,2.54$\,~\cite{Suzuki_JPSJ1995}, 
we obtain $H_{\rm c}\,{=}\,{\Delta}/g_{\parallel}{\mu}_{\rm B}\,{=}\,4.73$\,T, 
which is consistent with $H_{\rm c}\,{=}\,3.6$\,T evaluated in the present work.
From Eq.~(\ref{gap}), it is deduced that in CsFeCl$_3$, the application of pressure enhances 
the ratio of exchange interactions to the $D$ term. 
At $P$\,=\,$P_{\rm c}$ where the $\Delta$ value becomes zero, 
the condition $D$\,=\,2($2J_0$\,+\,3$J_1$) is satisfied.
According to the theoretical study in Ref.\,\onlinecite{Tsuneto_Physica1971},
the $H$\,$-$\,$T$ phase boundary of the ordered phase is determined only by the exchange interactions.
In addition, the temperature and field ranges are expected to be enhanced
with increasing exchange interactions.
These predictions are consistent with the obtained experimental results for CsFeCl$_3$.

In the isostructural compound CsFeBr$_3$, the intrachain exchange interaction 
is AF, in contrast to that in CsFeCl$_3$~\cite{TanakaY_JPSJ2001}. 
In CsFeBr$_3$, hydrostatic pressure increases the transition field $H_{\rm N}$ 
and decreases the transition temperature $T_{\rm N}$~\cite{Momosaki}, 
which indicates that the AF intrachain exchange interaction 
decreases with increasing applied pressure. 
This result is interpreted as follows.
Both AF and FM intrachain exchange paths are present in CsFeBr$_3$. 
The AF exchange interaction dominates the FM exchange interaction, 
and the resultant intrachain exchange interaction becomes AF. 
The FM exchange component increases with increasing pressure, 
thus, the resultant AF intrachain exchange decreases 
with increasing applied pressure. 
The magnitude of the AF exchange interaction $J_1$ in the $ab$-plane 
increases with increasing pressure because the lattice constant $a$ decreases. 
However, the effect of the pressure evolution of $J_1$ on the pressure-induced magnetic ordering 
is not considered to be dominant because $J_1$ is an order of magnitude smaller 
than the intrachain exchange interaction $J_0$.
We therefore deduce that the primary effect of pressure on this compound is to
enhance the FM intrachain exchange interaction $J_0$. 
Assuming that the values of $D$ and $J_1$ are unchanged under pressure, 
we estimate that $J_0/k_{\rm B}\,{=}\,5.91$\,K at the critical pressure 
$P_{\rm c}\,{\sim}\,0.9$ GPa, which is 1.12 times the value of $J_0/k_{\rm B}\,{=}\,5.27$\,K 
at ambient pressure\,~\cite{Yoshizawa_JPSJ1980}.

\section{Summary}

We have carried out low-temperature and high-pressure magnetization measurements 
on the gapped quantum magnet CsFeCl$_3$.
At ambient pressure, the $H$\,-\,$T$ phase diagram was determined down to 0.5\,K.
The exponent $\phi$ for the phase boundary of the field-induced ordered phase
decreases upon decreasing the analyzed temperature range.
It appears that $\phi$ converges to $\phi_{\rm BEC}$\,=\,1.5, 
the value predicted for three-dimensional magnon BEC. 
With increasing pressure, the phase boundary continuously extends to both
a lower-field and higher-temperature. 
The ground state was found to change at $P_{\rm c}$\,$\sim$\,0.9\,GPa 
from a gapped state to a magnetically ordered state in a very low applied field, 
indicating a pressure-induced quantum phase transition at $P_{\rm c}$ in CsFeCl$_3$.

Microscopic experiments such as electron spin resonance (ESR), 
nuclear magnetic resonance (NMR), and neutron scattering measurements
are of importance to determine the zero-field spin structure in the pressure-induced ordered phase
and the pressure dependence of the magnetic parameters.
In the pressure-induced ordered phase of CsFeCl$_3$, the massive Higgs mode might be observed
since the ordered moment is expected to be reduced 
by competition between the $D$ term and the exchange interactions.

\section*{Acknowledgments}
This work was supported by Grants-in-Aid for Scientific Research (A) Nos. 23244072 and 26247058, 
(C) No. 16K05414, and for Young Scientists (B) No. 26800181 from Japan Society for the Promotion of Science.


\begin{thebibliography}{25}
\bibitem{MPMS_BG}
The MPMS system evaluates the magnetization by 
analyzing the SQUID response as a function of sample position 
when the sample travels through a pickup coil.
Since the analysis assumes a point dipole moment,
it is difficult to estimate the background contribution
originating from a pressure device.
In this study, we obtained the magnetization of samples
by analyzing the difference signal 
between SQUID responses with and without samples
collected at each temperature and field.
\end{thebibliography}
\end{document}